# Large positive magnetoresistance in photocarrier doping potassium tantalites in the extreme quantum limit

Ruishu Yang, Dingbang Wang, Yang Zhao, Shuanhu Wang, Kexin Jin*

*Shaanxi Key Laboratory of Condensed Matter Structures and Properties and MOE Key Laboratory of Materials Physics and Chemistry under Extraordinary Conditions, School of Physical Science and Technology, Northwestern Polytechnical University, Xi'an 710072, China*

Abstract: We report on a high-field magnetotransport study of $KTaO_3$ single crystals. This material is a promising candidate to study in the extreme quantum limit (EQL). By photocarrier doping with 360 nm light, we have observed a significant positive, non-saturating, and linear magnetoresistance at low temperatures accompanied by a vanishing Hall coefficient. When cooled down to 10 K and subjected to a magnetic field of 12 T, the value of magnetoresistance of $KTaO_3$ (100) is increased by as much as 433%. Such behavior can be attributed to all electrons occupying only the lowest Landau level in the EQL. In this state, a quantum magnetoresistance is produced. This result provides novel insights into the next generation of magnetic devices based on complex materials and adds a new family of materials with positive magnetoresistance.

\_\_\_\_\_\_\_\_\_\_\_\_\_\_\_\_\_\_\_\_\_\_\_\_

[*] Corresponding author. E-mail addresses: jinkx@nwpu.edu.cn

Magnetoresistance (MR) has been extensively studied since its discovery in 1856. These materials are used in sensors, memory devices, and the emergent physics of complex interactions. MR effects have been investigated in magnetic and multilayer materials, such as ordinary MR [1, 2], anisotropic MR [3, 4], tunnel MR [5], giant MR [6], and colossal MR [7, 8]. Generally, the positive and conventional MR has a magnitude of only a few percent. It is negligible in some nonmagnetic materials due to the curving of carrier trajectory by the Lorentz force. A large and linear MR effect breaks this familiar rule. It has been observed in several non-magnetic semiconductors, such as Si [9], Bi [10, 11], $PtBi_2$ [12] and graphene [13]. Its discovery has raised a lot of lively discussions in the fields of condensed physics and functional materials. The kind of MR in nonmagnetic materials can avoid Barkhausen noise in magnetic recording and sensor devices.

Complex oxides have extraordinary and multifunctional properties. These include colossal MR in manganites, high-temperature superconductivity in cuprates, and multiferroics [14-19]. Among these materials, $KTaO_3$ (KTO) exhibits a cubic structure with a lattice constant of 0.3989 nm and a high dielectric constant (~ 4500). These characteristics make it a promising material for the extreme quantum limit (EQL) under a high magnetic field and at low temperatures [20]. Such a material could integrate its functions in the emerging electronics for all-oxide devices [21, 22]. The surface of KTO possesses a spectrum of emergent phenomena via ionic liquid gating, $Ar^+$ bombardment, and oxygen vacancy. These properties have been used in electrostatic superconductivity, topological Hall effect (THE), and de-Haas oscillation [23-26]. Recently, KTO has been used in the engineering design of two-dimensional electronic systems. Many heterointerfaces based on KTO have been studied, including EuO/KTO [27, 28], $LaTiO_3$/KTO [29], $LaVO_3$/KTO [30], amorphous-$LaAlO_3$/KTO [31], and $LaAlO_3$/KTO [32, 33]. They show intriguing properties, such as two-dimensional superconductivity, high-mobility spin-polarized electron gas, THE, and the anomalous Hall effect. So far, the MR in those systems is usually small, with a value less than 10% [34-37]. And a reliable large MR in complex oxides has not been found.

In this letter, we find that, under high magnetic fields (about 12 T), KTO single crystals with photocarrier doping come into the EQL state where the electrons are confined in the lowest Landau levels. A significant positive MR effect is achieved using 360 nm illumination with different light intensities at different temperatures. We observe that KTO exhibits a metallic behavior in the temperature range of 2-50 K. At these temperatures, their Hall coefficient vanishes as the magnetic field increases.

KTO (100) and (111) single crystals (3 × 3 mm) with different thicknesses are commercially available. The ultrasonically wire-bonded aluminum wires have electrical connections at their surface. Both the resistivity and the Hall effect were measured with a van de Pauw geometry under 360 nm illumination with different light intensities and thicknesses. We used a Physical Property Measurement System (CFMS-14T) that was equipped with an optical fiber. This apparatus is shown in the inset of Fig. 1(a). The bandgap ($E_g$) of KTO is 3.4 eV, and the wavelength of light is 360 nm (3.44 eV). The electrons can be promoted from the valence band to the conduction band under 360 nm light. Although the light illumination generates non-equilibrium carriers, a steady-state quasi-equilibrium can be achieved because the carrier lifetime is close to 0.15 ms [38]. This lifetime is much larger than thermalization. The holes are straightforward to trap, and thus the transport is dominated by electrons [20, 38, 39]. Furthermore, Hall measurements confirm that electrons mainly determine the conduction. More importantly, the interaction between light and matter depends on the optical properties of matter and the wavelength of light. The optical penetration depth is deduced from the absorption coefficient ($\alpha$) of KTO, from which the depth profile of electronic distribution can be estimated. From the absorption coefficient, we can obtain that the thickness of conductive layer is larger than 1 mm at $\lambda$ = 360 nm. At this wavelength, the carrier density distribution, which is perpendicular to the surface of samples, can be neglected. This distance significantly exceeds the thickness of KTO crystals. Further, we define MR=[$\rho_{xx}(H)$–$\rho_{xx}(H=0)$]×100%/$\rho_{xx}(H=0)$. Supplementary Fig. S1 shows the relationship between MR and $H$ when the magnetic field is perpendicular and parallel to the current at 20 K. That data indicates that the samples exhibit isotropic properties. Considered with the thickness of the conductive layer, the electric conduction of KTO under 360 nm light is quasi-three-dimensional. Additionally, LaAlO$_3$/KTO heterointerfaces were prepared using pulsed laser deposition at 800 °C and 1×10$^{-3}$ Pa of O$_2$ with a KrF excimer laser ($\lambda$ = 248 nm) that operated at 1 Hz.

The temperature dependences of the electrical resistivity of KTO (100) and (111) are shown in Fig. 1(a). The inset shows the measurements of resistivity and Hall resistivity. We observe that the KTOs show a metallic behavior at low temperatures. The temperature dependence of the electrical resistivity typically can be described by a power law, $\rho=\rho_0+AT^n$ [40]. When $n = 2$, the expected behavior is a Fermi liquid, while a 'non-Fermi liquid' is identified with $1 < n < 2$. The fitted values for $n$ are 1.62 for KTO (100) and 1.64 for (111) at low temperatures. These values indicate that they are non-Fermi liquid.

To analyze the electronic transport of KTO (100) by photodoping, we measure its Hall resistivity ($\rho_{xy}$) under the magnetic field ($H$). Fig. 1(b) shows the behavior at different temperatures. Strikingly, the $\rho_{xy}$–$H$ is nonlinear below 20 K. The carrier density ($n$) and mobility ($\mu$) can be obtained from the Hall coefficient at a zero magnetic field. The carrier density slightly decreases as the temperature decreases, while the mobility increases and reaches 1200 cm$^2$/Vs at 2 K. More importantly, the KTOs still exhibit a metallic behavior even at very low electron density $n = 1.4\times10^{12}$ cm$^{-3}$. We also measured the carrier density and mobility of KTO single crystals with various light intensities at 2 K. Supplementary Fig. S2 demonstrates that the carrier density increases with the light intensity, which is consistent with previous research of STO single crystals [41].

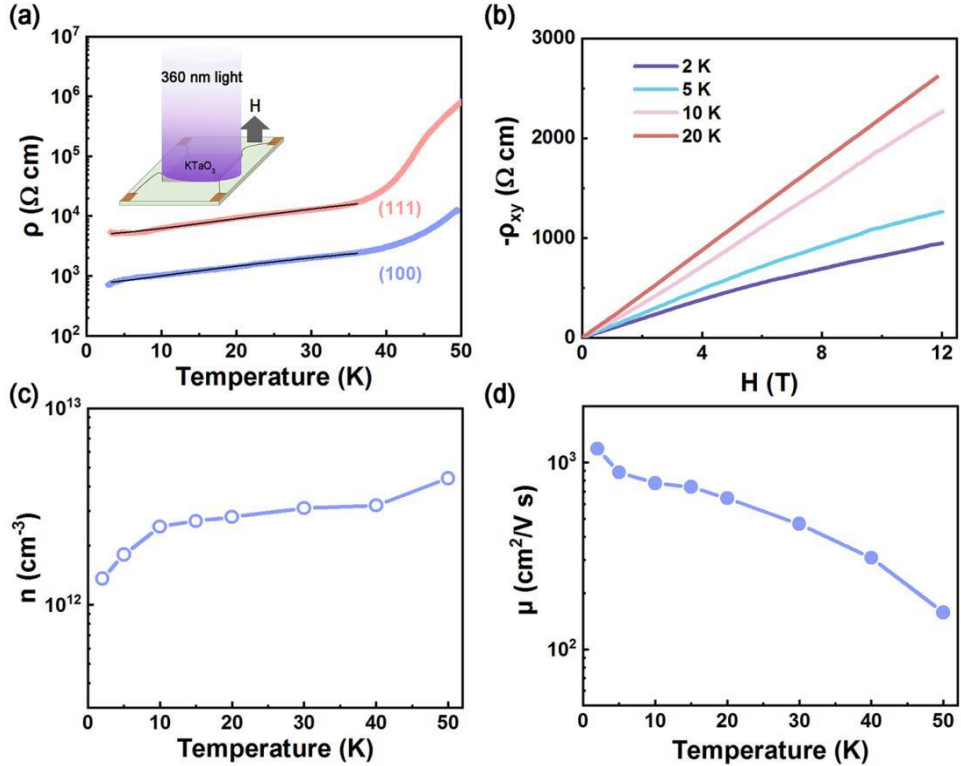

FIG. 1. (a) The temperature dependence of the resistivity of KTO (100) and (111) under the illumination at λ=360 nm and 33 mW/cm$^2$. Inset of (a) shows a sketch for the measurements of resistivity and Hall resistivity. (b) Hall resistivity ($\rho_{xy}$) of KTO (100) under the illumination as a function of magnetic field at different temperatures. (c) Carrier density $n$ and (d) mobility $\mu$ of KTO (100) as a function of temperature.

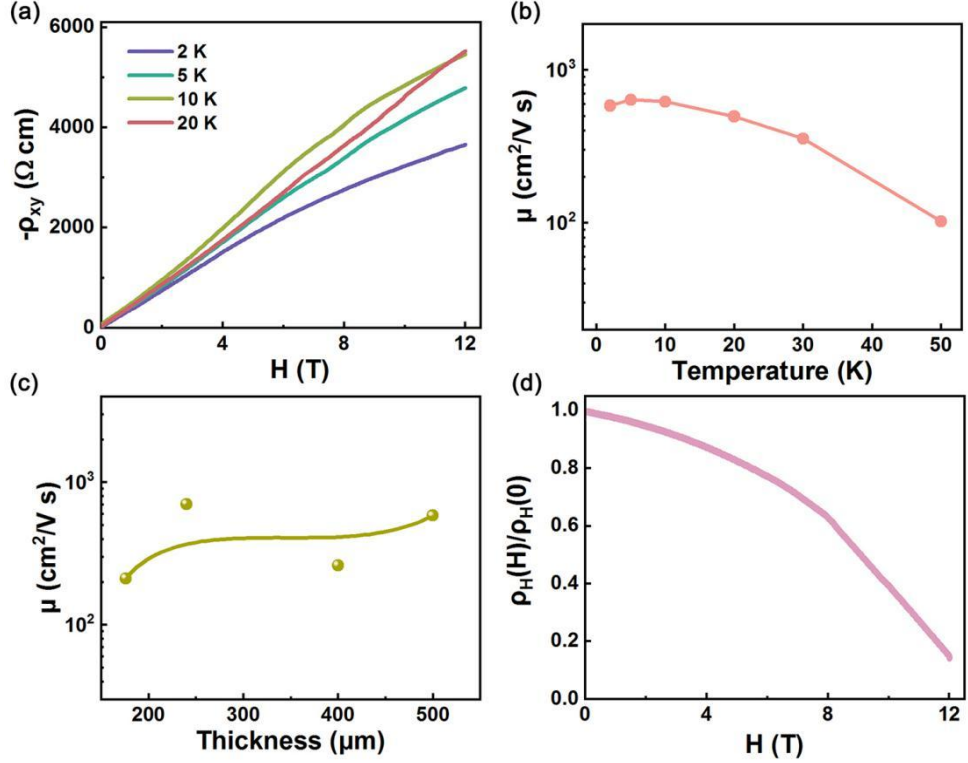

FIG. 2. (a) Hall resistivity ($\rho_{xy}$) of KTO (111) under the illumination in the temperature range of 2-20 K. (b) and (c) are the mobility of KTO (111) as a function of temperatures and thicknesses at 0 T under the illumination with $\lambda$=360 nm and 33 mW/cm$^2$, respectively. (d) The normalized Hall coefficient of KTO under illumination as a function of magnetic field at 2 K.

The Hall resistivity of KTO (111) in the temperature range of 2-20 K is shown in Fig. 2(a). The Hall resistivity ($\rho_{xy}$) favors a nonlinear dependence at low temperatures. The mobilities of KTO (111), as a function of thicknesses and temperatures, are shown in Fig. 2(b) and 2(c), respectively. The mobility is enhanced as the temperature decreases and the thicknesses increase. Generally, the nonlinear Hall effect has two origins: one is the abnormal Hall effect caused by magnetism, and the other is caused by the coexistence of two or more types of carriers. The abnormal Hall effect can be determined by the derivative relationship between $\rho_{xy}$ and $H$ if the derivation curve has a peak near 0 T [42]. We exclude the abnormal Hall effect according to the curves of $d\rho_{xy}/dH$ (Supplementary Fig. S3). Furthermore, the multiband conduction or parallel conduction channels also is ruled out (Supplementary Fig. S4 and Table S1).

The relationship between the magnetic field and the normalized Hall coefficient is shown in Fig. 2(d) to investigate this anomalous feature further. The normalized Hall coefficient decreases with increasing magnetic field, consistent with reports by Kozuka *et al.* [43]. The vanishing of the Hall coefficient is probably caused by the low level of scattering under a high magnetic field for the EQL.

The requirements to reach EQL are $\omega_c\tau > 1$ and $\hbar\omega_c > k_BT, E_F$, where $\omega_c = eH/m^*$ is the cyclotron frequency ($e$ is the elementary charge, $H$ is the magnetic field, and $m^*$ is the electron effective mass), $\tau$ is the carrier relaxation time, $k_BT$ is the thermal energy, and $E_F$ is the Fermi energy. In this work, $\hbar\omega_c = 2.78$ $meV \gg k_BT = 0.18$ $meV$ and $E_F = \hbar^2k_F^2/2m^* = 5.4\times10^{-4}$ $meV$. In addition, applying the Mott criterion to $\hbar\omega_c > k_BT, E_F$, the EQL can be reached by satisfying the following condition: $\varepsilon_r/m_r > 1.3 \times 10^3$, where $\varepsilon_r = \varepsilon/\varepsilon_0$ ($\varepsilon_0$ is the vacuum permittivity) and $m_r = m^*/m_0$ ($m_0$ is the bare electron mass). For the KTOs in this work, the value of $\omega_c\tau$ is ~1.44, larger than 1 at a high magnetic field because of the high mobility (>$10^3$ cm$^2$/Vs). Further, the dielectric constant ($\varepsilon_r$) of KTO is ~4500 (2 K) [20] and $m^*$ is $0.8m_0$ [20, 39]. Therefore, $\varepsilon_r/m_r$ of ~$5.6\times10^3$ is larger than $1.3\times10^3$. The condition should also be satisfied. Therefore, KTO comes into an EQL state. Electronic wave functions are highly localized in the EQL. The small gyration radius of electrons in the perpendicular direction reduces the probability of scattering between electrons. Hence, the Hall coefficient decreases as the magnetic field increases.

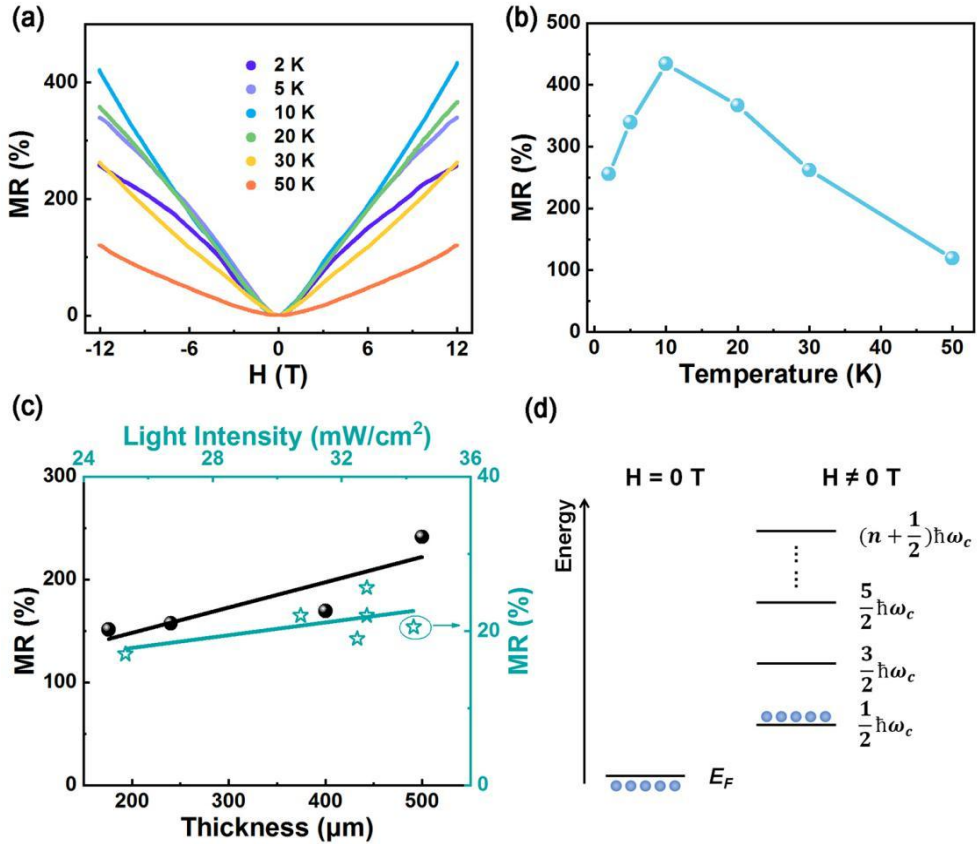

FIG. 3. (a) MR of KTO (100) as a function of magnetic field at different temperatures. (b) MR of as a function of temperatures with KTO (100) at 12 T. (c) The MR of KTO (111) dependent on different thicknesses (12 T) and light intensities (2 T) at 2 K, the solid lines are linear fitting curves. (d) Schematic diagram of the energy level under $H = 0$ T and $H \neq 0$ T. All electrons occupy at only the lowest Landau level and other levels are empty when $\hbar\omega_c \gg E_F$.

Fig. 3(a) shows the relationship between the MR of KTO (100) and the magnetic field. The result for KTO (111) is shown in Supplementary Fig. S5. The MR displays giant positive, linear, and non-saturating features. As shown in Fig. 3(b), the MR increases nonlinearly with decreasing temperature. The MR values with a thickness of 500 μm under 12 T are 256%, 340%, 433%, 367%, 263%, and 119% at 2 K, 5 K, 10 K, 20 K, 30 K, and 50 K, respectively. These values are far more significant than the ordinary MR effect. The maximum value might be attributed to the localization of electrons below 10 K due to the ferroelectric phase of KTO [44]. This phenomenon is consistent with the drop in carrier density. Furthermore, we measured the MR of KTOs with different thicknesses and light intensities at 2 K. Fig. 3(c) shows that the MR of KTO (111) reaches ~200% (12 T) with different thicknesses and ~20% (2 T) with various light intensities (25-35 mW/cm$^2$). These values are far greater than that of oxide interfaces [24, 29].

Early in 1959, Lifshits and Peschansky [45] proposed that the linear MR effect could be caused when the Larmor radius of electrons was smaller than their mean free path for a metal in a high magnetic field. Further, Abrikosov declared [46, 47] that the system would reach EQL for a material with a small $m^*$ if the magnetic field was very high. This conjecture assumes a gapless semiconductor with a linear energy spectrum and that all of the electrons would occupy only the lowest Landau energy level. This condition produces a linear MR with a magnetic field, $\rho_{xx} \propto H$. Afterward, various other theories have been proposed to explain the positive MR in nonmagnetic materials by mechanisms that include electric field inhomogeneity [48], density inhomogeneity [49], density fluctuations [50], and antiferromagnetic fluctuations [51]. In our work, although the carrier density is very low, the photocarrier-doped KTO favors metallic at low temperatures because a combination of a small $m^*$ and a large dielectric constant can push the metal-insulator transition boundary to low densities [9]. The electrons are distributed below the Fermi surface when $H = 0$ T, as shown in Fig. 3(d), and $E_F$ decreases as increasing $H$. At such a low carrier density and high magnetic field, the Fermi energy of $E_F = 5.4 \times 10^{-4}$ *meV* is very low. Thus the lowest Landau energy, $\hbar\omega_c = 2.78$ *meV,* is far larger than $E_F$. At this time, the electrons occupy the lowest Landau level, and all other levels are empty. Fig. 3(d) illustrates this state. As a result, KTO shows a quantum MR effect [13, 46, 52]. A non-saturating MR further verifies that the system of photocarrier doped KTO comes into the EQL. Another interpretation is the inhomogeneity in bulk or films [48-51], which might arise from the carrier density inhomogeneity, mainly caused by the light distribution from an uneven surface. To further analyze our experimental results, we measured the MR of STO single crystals and $LaAlO_3$/KTO heterointerfaces under 360 nm light illumination. Supplementary Fig. S6 shows no significant and linear MR effect. Therefore, the carrier density inhomogeneity of light distribution can be ruled out.

In summary, KTO single crystals exposed to 360 nm light show a metallic behavior at low temperatures with low carrier density and high mobility. We conclude that these photodoping KTOs reach EQL state, which therefore provide a new oxide material with EQL. Further, we have discovered a significant positive MR effect and the vanishing of the Hall coefficient at low temperature and a high magnetic field. In particular, the MR value of KTO (100) reaches 433% at 10 K under 12 T. This state is caused by all electrons occupying the lowest Landau level at EQL. This work paves a way to understand the fundamental physics of the interaction between light and complex materials.

## Acknowledgments

This work is supported by the National Natural Science Foundation of China (Nos. 51572222), Key Research Project of the Natural Science Foundation of Shaanxi Province, China (Grant No. 2021JZ-08 and 2020JM-088), the Natural Science Basic Research Plan in Shaanxi Province of China (2021JM-041), and the Fundamental Research Funds for the Central Universities (3102017OQD074, 310201911cx044).

## Data Availability

The data that support the findings of this study are available from the corresponding author upon reasonable request.

# Supplemental Material

# Large positive magnetoresistance in photocarrier doping potassium tantalites in the extreme quantum limit

Ruishu Yang, Dingbang Wang, Yang Zhao, Shuanhu Wang, Kexin Jin [*]

*Shaanxi Key Laboratory of Condensed Matter Structures and Properties and MOE Key Laboratory of Materials Physics and Chemistry under Extraordinary Conditions, School of Physical Science and Technology, Northwestern Polytechnical University, Xi'an 710072, China*

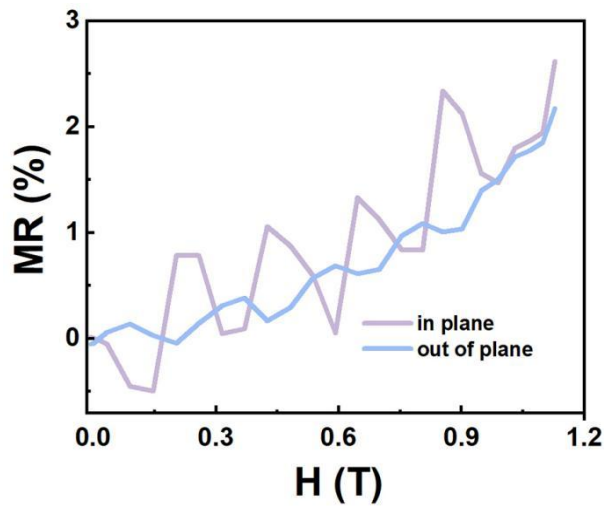

Figure S1. MR as a function of perpendicular (out of plane) and parallel (in plane) magnetic field measured at 20 K.

---

[*] Author to whom correspondence should be addressed. E-mail addresses: jinkx@nwpu.edu.cn

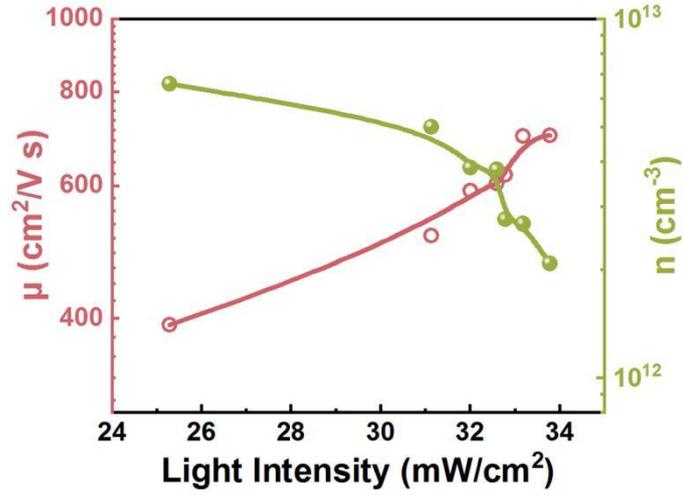

Figure S2. Mobility (μ) and carrier density (n) of KTaO$_3$ (111) as a function of light intensity at 2 K.

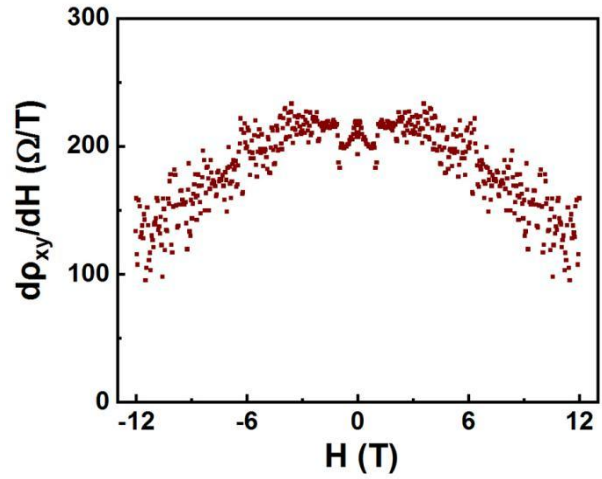

Figure S3. Derivative of hall resistivity with respect to magnetic field at 2 K.

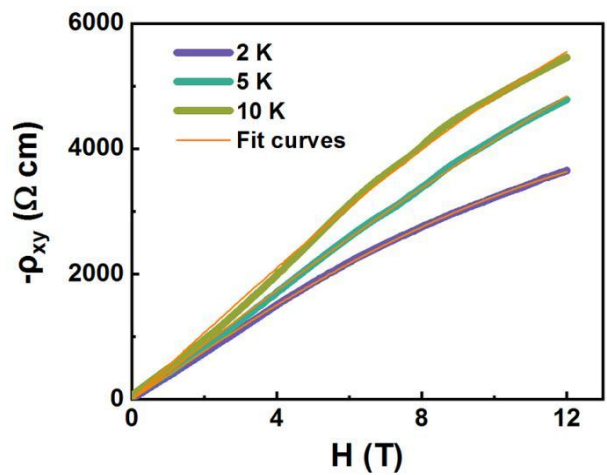

Figure S4. Hall resistivity ($\rho_{xy}$) as a function of magnetic field under illumination at 2-10 K, fit curves are using the two-channel conduction from the electronic bands.

Table S1 Fitting parameters of the $\rho_{xy}$ vs. $H$ by using Eq. 1.

| Parameter T | $n_1$ (cm$^{-3}$) | $\mu_1$ (cm$^2$/V s) | $n_2$ (cm$^{-3}$) | $\mu_2$ (cm$^2$/V s) |
|---|---|---|---|---|
| 2 K | 7.75E12 | -19.4 | 4.25E11 | 2251.8 |
| 5 K | 2.79E11 | -76.4 | 2.55E11 | 1506.4 |
| 10 K | 5.01E12 | -8.2 | 1.68E11 | 1123.7 |

We fitting $\rho_{xy}$-$H$ used the two-band conduction, which is given by the following formula:

$$\rho_{xy} = \frac{H(n_1\mu_1^2+n_2\mu_2^2)+H(\mu_1\mu_2 H)^2(n_1+n_2)}{e(n_1\mu_1+n_2\mu_2)^2+e(\mu_1\mu_2 H)^2(n_1+n_2)^2} \quad (1)$$

the results by fitting show negative values (listed in the table), which is unreasonable. Thence, we ruled out the possibility of two types carriers.

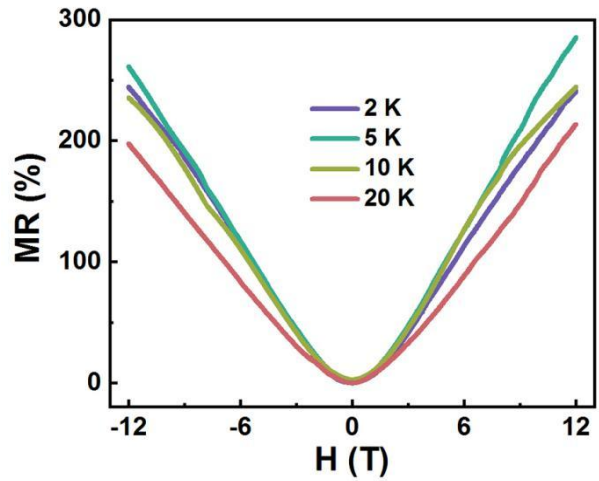

Figure S5. MR of KTaO$_3$ (111) as a function of magnetic field at different temperatures.

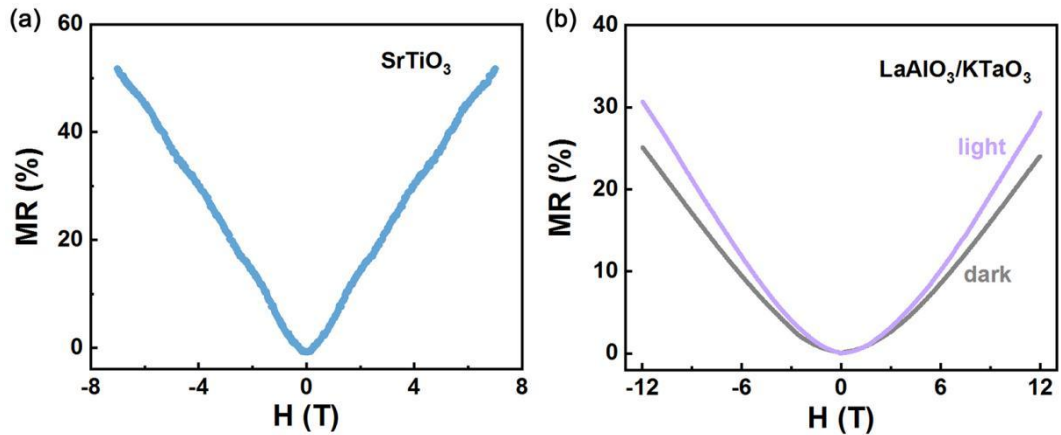

Figure S6. (a) MR of SrTiO$_3$ (100) as a function of magnetic field at 2 K. (b) MR of LaAlO$_3$/KTaO$_3$ (100) in dark and 360 nm illumination at 2 K.